\documentclass[a4paper,11pt]{article}
\usepackage{jinstpub} 
\usepackage{lineno}
\usepackage{enumitem}
\usepackage{subcaption}

\title{\boldmath Characterization of afterpulse in SiPMs with single-cell readout as a function of bias voltage and fluence}

\author[a,1]{P. Parygin,\note{Corresponding author.}}
\author[a]{E. Garutti,}
\author[b]{E. Popova,}
\author[a]{and J. Schwandt}

\affiliation[a]{Universität Hamburg, Institut für Experimentalphysik,\\
Luruper Chaussee 149, 22761 Hamburg, Germany}
\affiliation[b]{University of Maryland,\\
College Park, MD 20742, United States}

\emailAdd{pavel.parygin@cern.ch}

\abstract{We present a detailed investigation of the afterpulse effect in silicon photomultipliers (SiPMs), using a dedicated structure with single-cell readout, which enables direct measurement of intrinsic device properties and observation of individual pulses also after irradiation.

Three independent analysis methods to quantify afterpulse induced events were developed and validated by Monte Carlo simulations. The first method is based on charge integration, while the other two methods use multiple linear regression to reconstruct transient waveforms and accurately identify individual pulse positions. These pulse positions are then used either as direct event counts or to construct time interval distributions, enabling comprehensive characterization of the afterpulse probability and providing insights into the dynamics of trapping in silicon.

Using this framework, we measured three SiPM samples with single-cell readout: one fresh reference device and two irradiated devices that were exposed to reactor neutron fluences of \mbox{$\Phi = 2\cdot10^{12}$, $1\cdot10^{13}$~cm$^{-2}$}.

We report systematic measurements of the afterpulse probability and time constant as functions of bias voltage and irradiation fluence. For overvoltages in the range of 3–5 V above breakdown, the afterpulse time constant is found to be below 10 ns and the afterpulse probability below $6~\%$. Both quantities show no significant dependence on irradiation fluence.}

\keywords{Silicon photomultiplier, single cell SiPM, radiation damage, afterpulse, characterization}

\arxivnumber{2604.08308} 

\begin{document}
\maketitle
\flushbottom

\section{Introduction}
\label{sec:intro}

Silicon Photomultipliers (SiPMs) have transitioned from experimental prototypes to a baseline technology for high-energy physics (HEP) \cite{Musienko2015} due to their high gain, single-photon resolution, and magnetic field immunity. However, performance is constrained by stochastic noise, categorized into primary uncorrelated events — dominated by thermal and tunneling-induced Dark Count Rate (\textit{DCR}) — and secondary correlated phenomena. The latter includes optical crosstalk (\textit{CT}) and afterpulsing (\textit{AP}). While \textit{CT} involves neighboring cells, \textit{AP} is a temporal distortion within the primary cell caused by carrier trapping and subsequent release, or by delayed diffusion of minority carriers from the non-depleted bulk.

In standard arrays, the temporal overlap of these phenomena complicates the extraction of pure trap parameters \cite{Klanner2019}. To decouple these effects, this study utilizes a dedicated SiPM structure enabling independent biasing and readout of a single microcell. By isolating the cell, spatial crosstalk is suppressed, allowing for precise characterization of the temporal distribution of secondary discharges and the determination of intrinsic trap lifetimes.

This work systematically investigates afterpulsing dynamics across overvoltages ($\Delta U$) up to 5~V and radiation fluences up to $\Phi = 1\cdot10^{13}$~cm$^{-2}$. By analyzing the temporal distribution of secondary events within a $600$~ns window from the primary pulse, we characterize the underlying trap properties, specifically the de-trapping time constants ($\tau_{AP}$) and total afterpulse probability ($P_{AP}$).
\section{Device and experimental setup}
\label{sec:device_setup_data}

The measurements utilized a dedicated single-cell SiPM (Hamamatsu S14160) consisting of an $11 \times 11$ pixel array with a $15~\mu$m pitch. A central pixel is physically decoupled with a dedicated output for independent biasing and readout; its structure and baseline performance are detailed in \cite{BYCHKOVA2022166533, BYCHKOVA2022167042}. This study analyzes a non-irradiated reference and two samples irradiated to neutron fluences of $\Phi = 2\cdot10^{12}$ and $10^{13}~$cm$^{-2}$. Raw waveforms were captured at 10~GS/s within a $1~\mu$s window, synchronized to a 100~kHz, 451~nm laser trigger at $\approx 310$~ns. The central cell was biased at $U_{bd} + [2, 5]$~V, while surrounding pixels remained slightly below $U_{bd}$ to minimize interference. The laser intensity was attenuated to 50~\%. This configuration enables the precise isolation of primary discharges and subsequent secondary pulses, which is necessary for characterizing afterpulsing dynamics.
\section{Data analysis method}
\label{sec:analysis}

\begin{figure}[htbp]
    \centering
    \begin{subfigure}[t]{0.48\textwidth}
        \centering
        \includegraphics[width=\textwidth, trim={0 0 0 0.8cm}, clip]{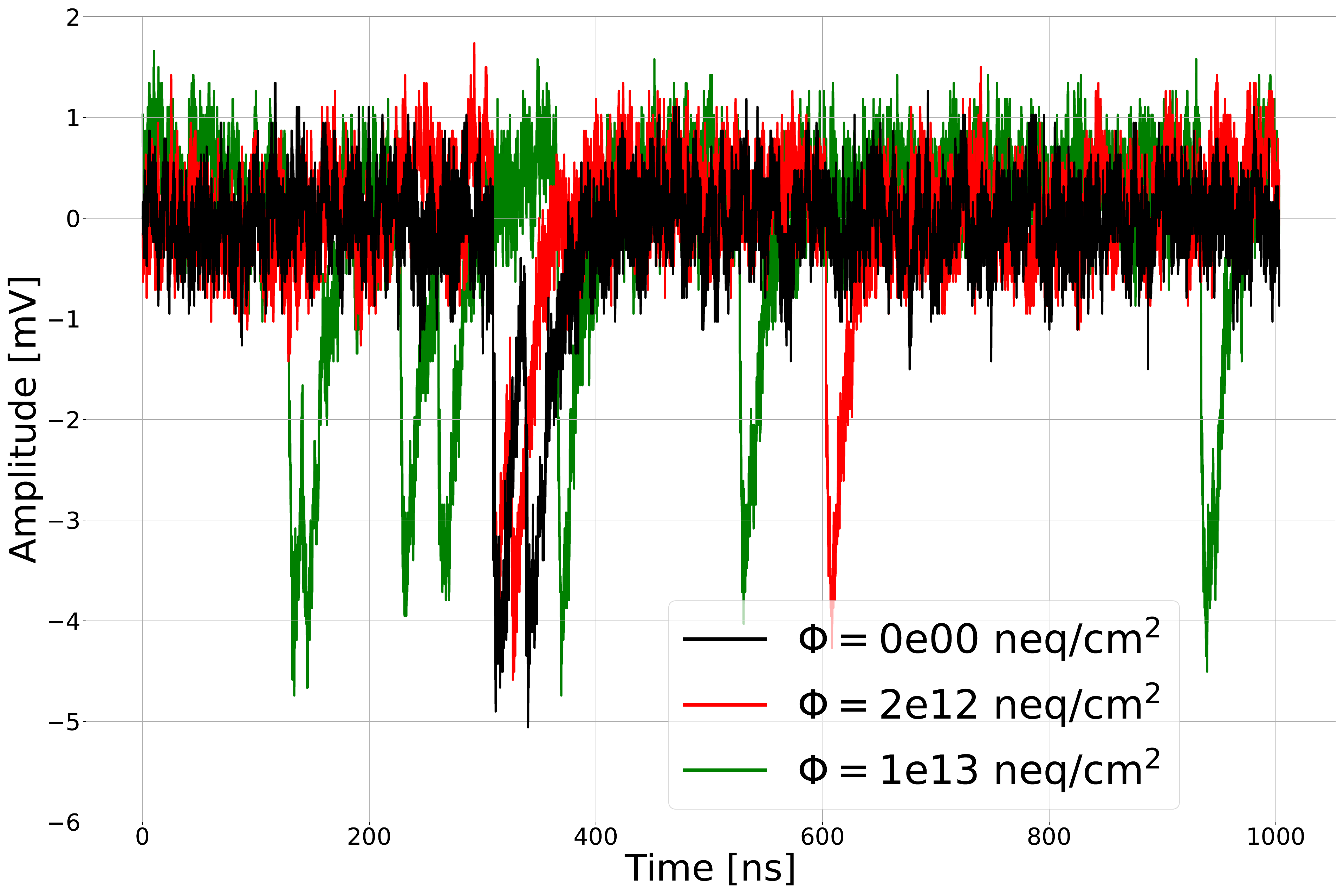}
        \caption{Experimental data.}
        \label{fig:raw_wf}
    \end{subfigure}
    \hfill
    \begin{subfigure}[t]{0.48\textwidth}
        \centering
        \includegraphics[width=\textwidth, trim={0 0 0 0.8cm}, clip]{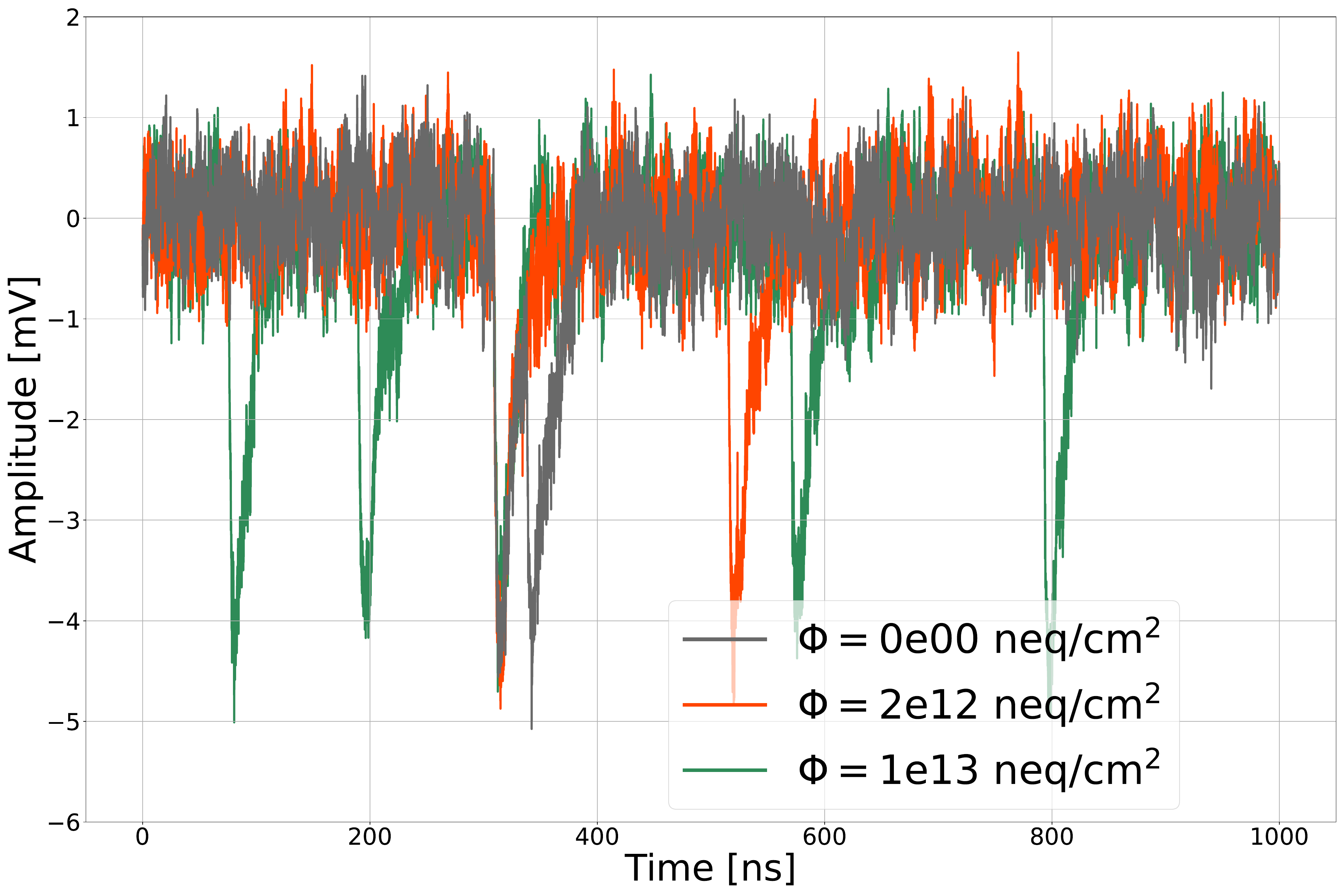}
        \caption{Monte Carlo simulation.}
        \label{fig:raw_wf_sim}
    \end{subfigure}
    \caption{SiPM response waveforms for $\Phi = 0$ (black), $2\cdot10^{12}$ (red), and $1\cdot10^{13}~$cm$^{-2}$ (green).}
    \label{fig:pic_raw_waveforms}
\end{figure}

The extraction of afterpulsing parameters follows a three-stage automated pipeline designed to handle the waveforms from the single-cell device. Examples of raw waveforms for different fluences are shown in figure~\ref{fig:raw_wf}.

\subsection{Pulse Identification}
\label{sec:pulse_finder}
\begin{figure}[htbp]
     \centering
     \begin{subfigure}[t]{0.32\textwidth}
         \centering
         \includegraphics[width=\textwidth, trim={0 0 0 0.5cm}, clip]{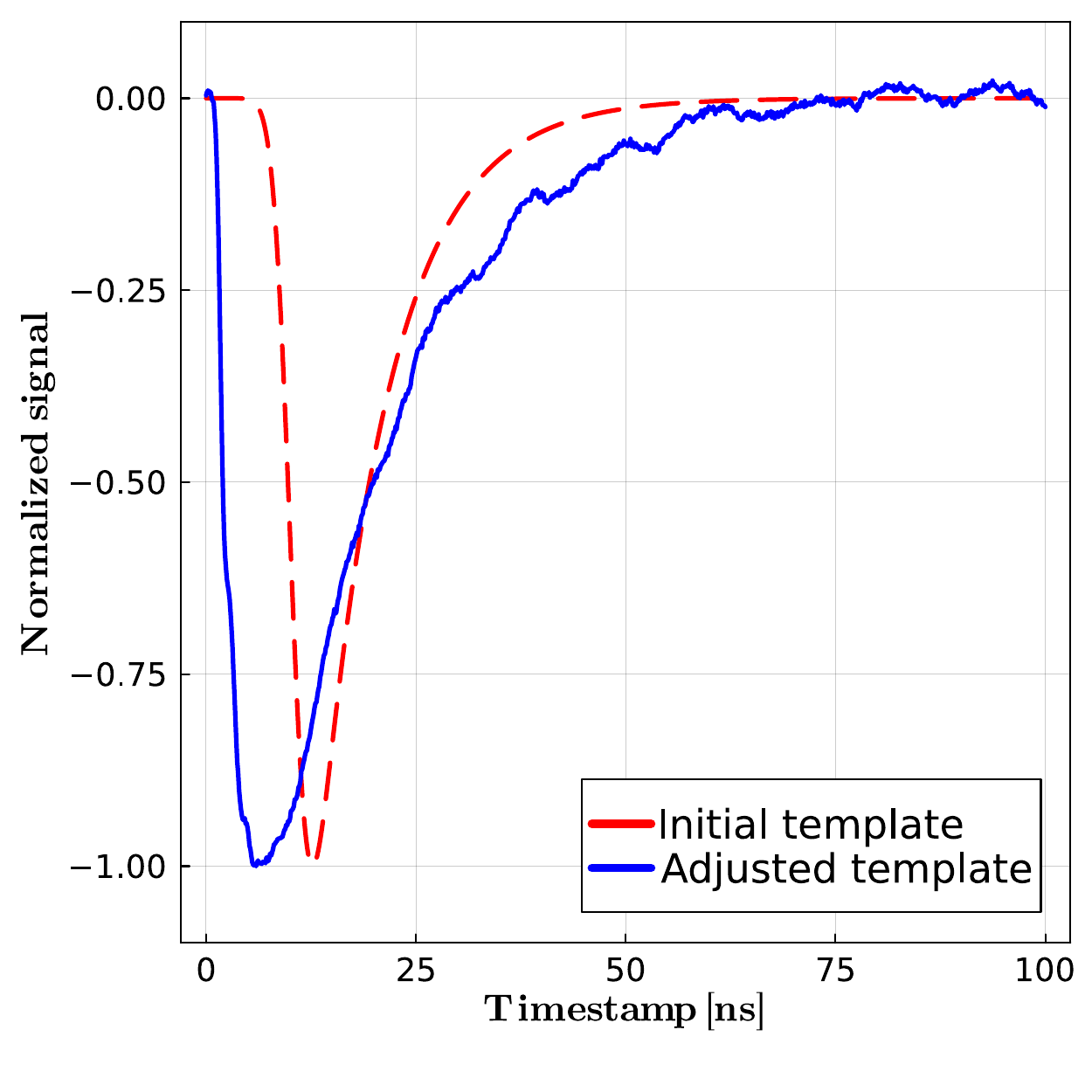}
         \caption{Initial (red) and adjusted (blue) templates.}
         \label{fig:template}
     \end{subfigure}
     \hfill
     \begin{subfigure}[t]{0.32\textwidth}
         \centering
         \includegraphics[width=\textwidth, trim={0 0 0 0.5cm}, clip]{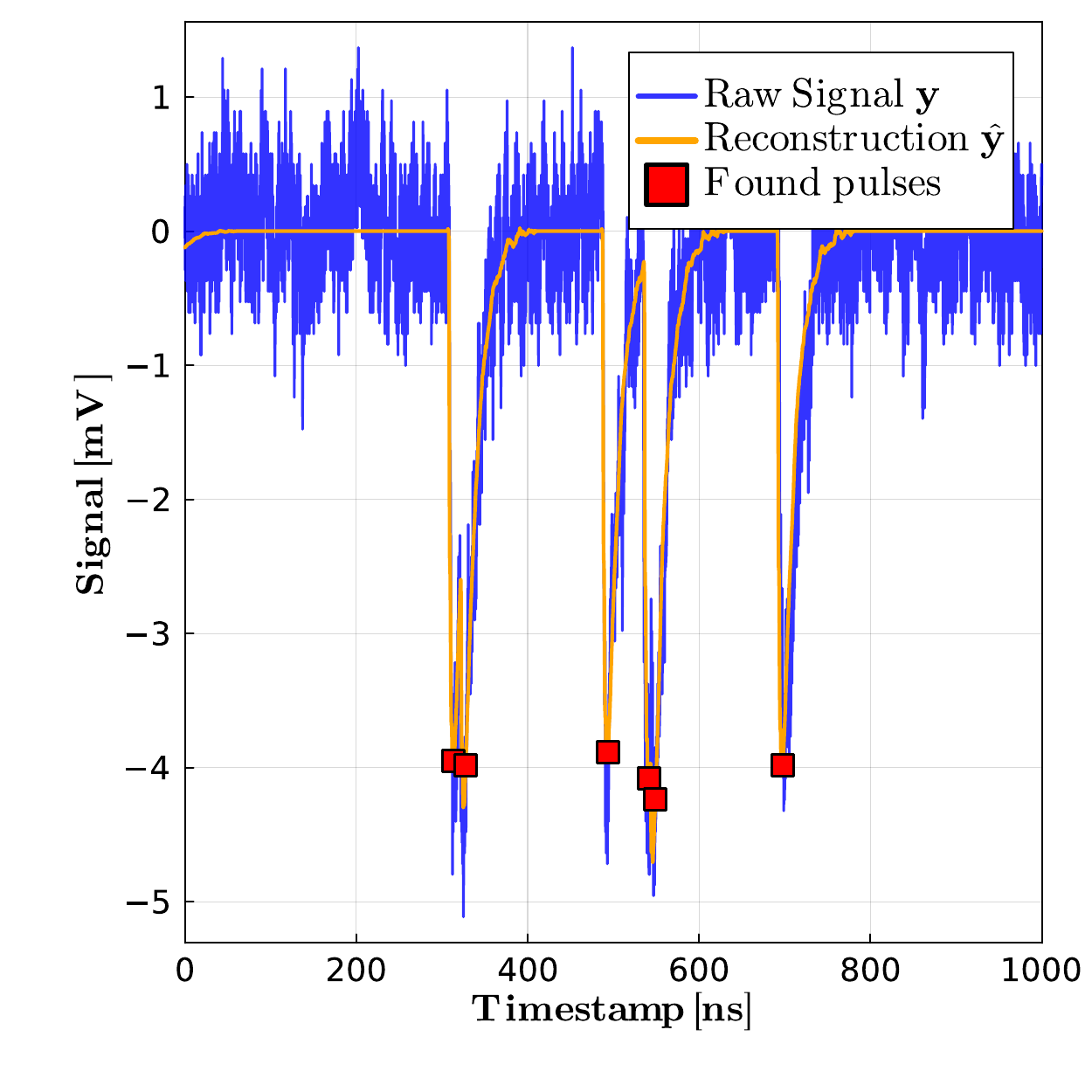}
         \caption{Pulse reconstruction at \mbox{$\Phi=1\cdot10^{13}$~cm$^{-2}$}.}
         \label{fig:reconstruction}
     \end{subfigure}
     \hfill
     \begin{subfigure}[t]{0.32\textwidth}
         \centering
         \includegraphics[width=\textwidth, trim={0 0 0 0.5cm}, clip]{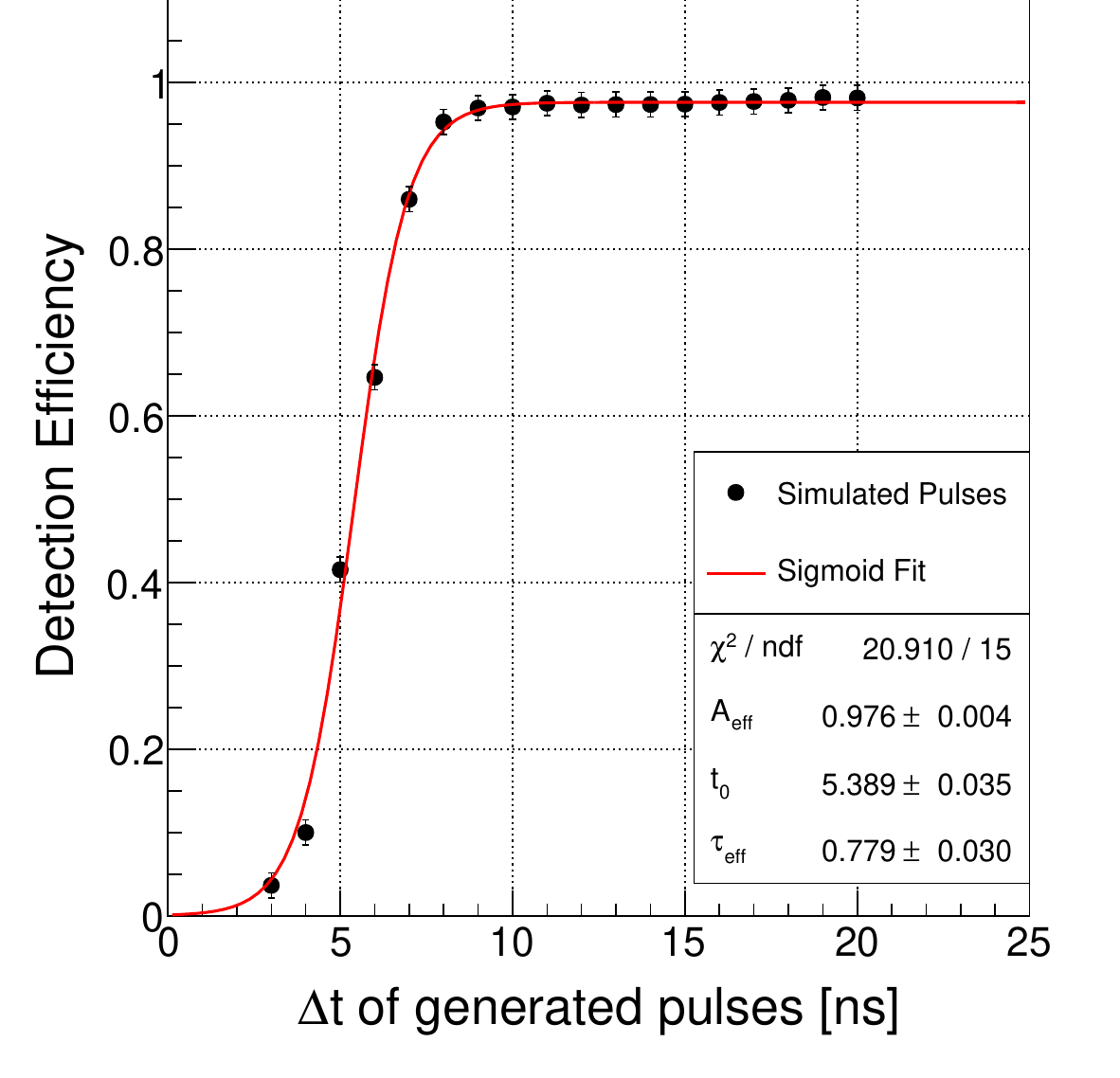}
         \caption{PF efficiency versus pulse separation $\Delta t$.}
         \label{fig:eff}
     \end{subfigure}
     \caption{Pulse finding (PF) algorithm performance. (a) Normalized single-photon responses. (b) Raw signal $y$ (blue), reconstructed signal $\hat{y}$ (orange), and identified pulses (red squares). (c) Sigmoid fit (red line) showing a 50\% efficiency threshold at $t_0 \approx 5$~ns.}
     \label{fig:algo}
\end{figure}

To accurately identify pulses in high-rate or overlapping scenarios, we utilize a pulse-finding (PF) algorithm based on Multiple Linear Regression (MLR) \cite{Schmailzl_2023}. The raw waveform $y(t)$ is modeled as:
\begin{equation}
\hat{y}(t) = \sum_{i=1}^{n} A_i \cdot T(t - t_i) + B,
\end{equation}
where $A_i$ and $t_i$ are the amplitude and arrival time of the $i$-th pulse, $T$ is the average single-cell response (Fig.~\ref{fig:template}), and $B$ is the dynamic baseline. Minimizing the residual sum of squares allows the algorithm to discriminate consecutive discharges "hidden" on the falling edge of primary pulses (Fig.~\ref{fig:reconstruction}). 

The algorithm efficiency $\epsilon$ is characterized via Monte Carlo trials (Sec.~\ref{sec:simulations}) and modeled by a sigmoid function  $\epsilon(\Delta t) = A_{eff} / \left(1 + e^{ -(\Delta t - t_0) / \tau_{eff}}\right)$ (Fig.~\ref{fig:eff}).
Here, $A_{eff}$ is the plateau efficiency, $t_0$ is the 50\% efficiency threshold, and $\tau_{eff}$ defines the turn-on slope. Electronic noise prevents $A_{eff}$ from reaching 100\% as it occasionally distorts pulses beyond the template-matching threshold.

\subsection{DCR Calculation and Event Selection}
\begin{figure}[htbp]
    \centering
    \begin{subfigure}[t]{0.42\textwidth}
        \centering
        \includegraphics[width=\textwidth, trim={0 0 0 0.4cm}, clip]{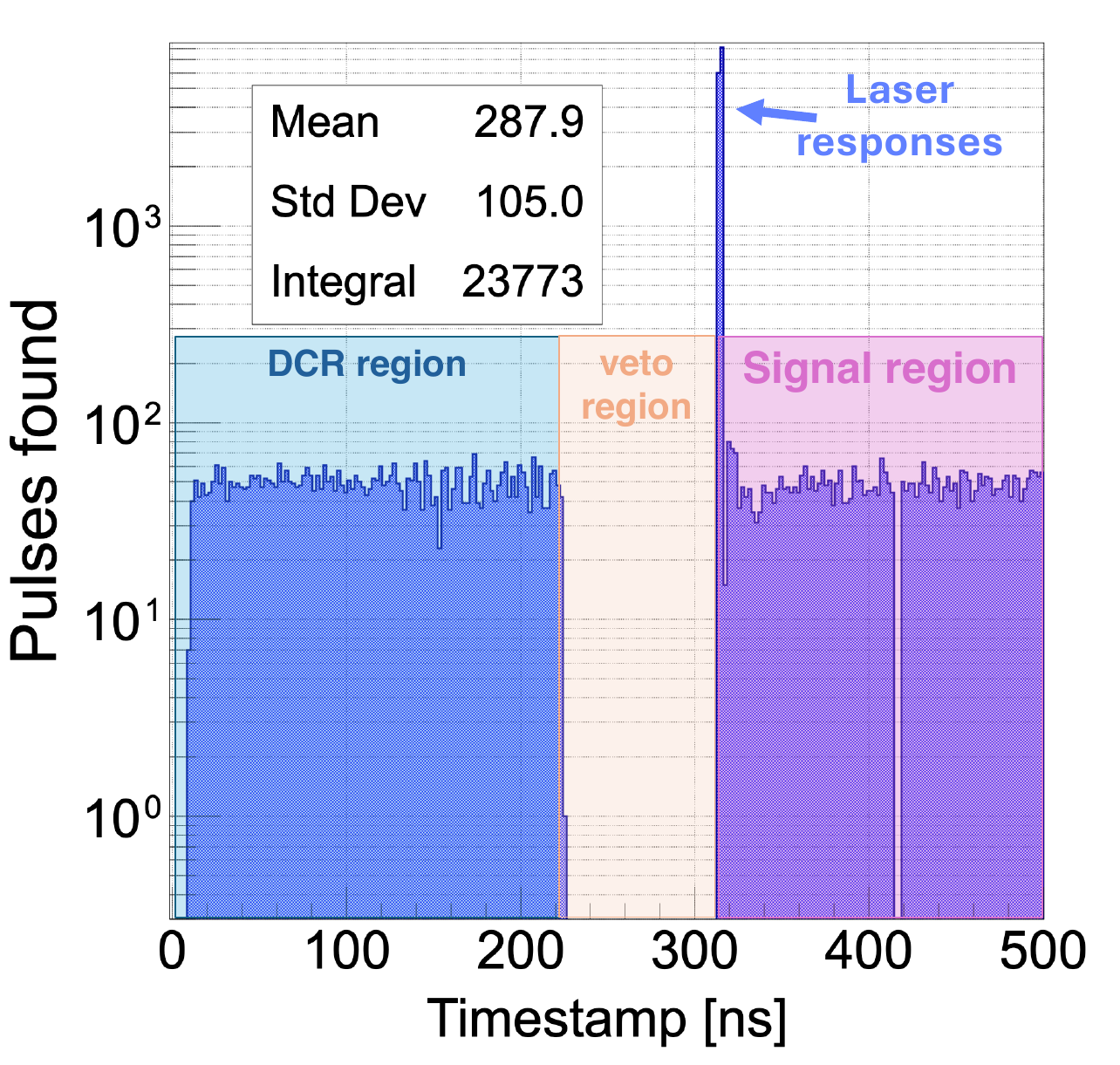}
        \caption{Detected pulse distribution.}
        \label{fig:found_pulses}
    \end{subfigure}
    \hfill
    \begin{subfigure}[t]{0.54\textwidth}
        \centering
        \includegraphics[width=\textwidth, trim={0 0 0 0.4cm}, clip]{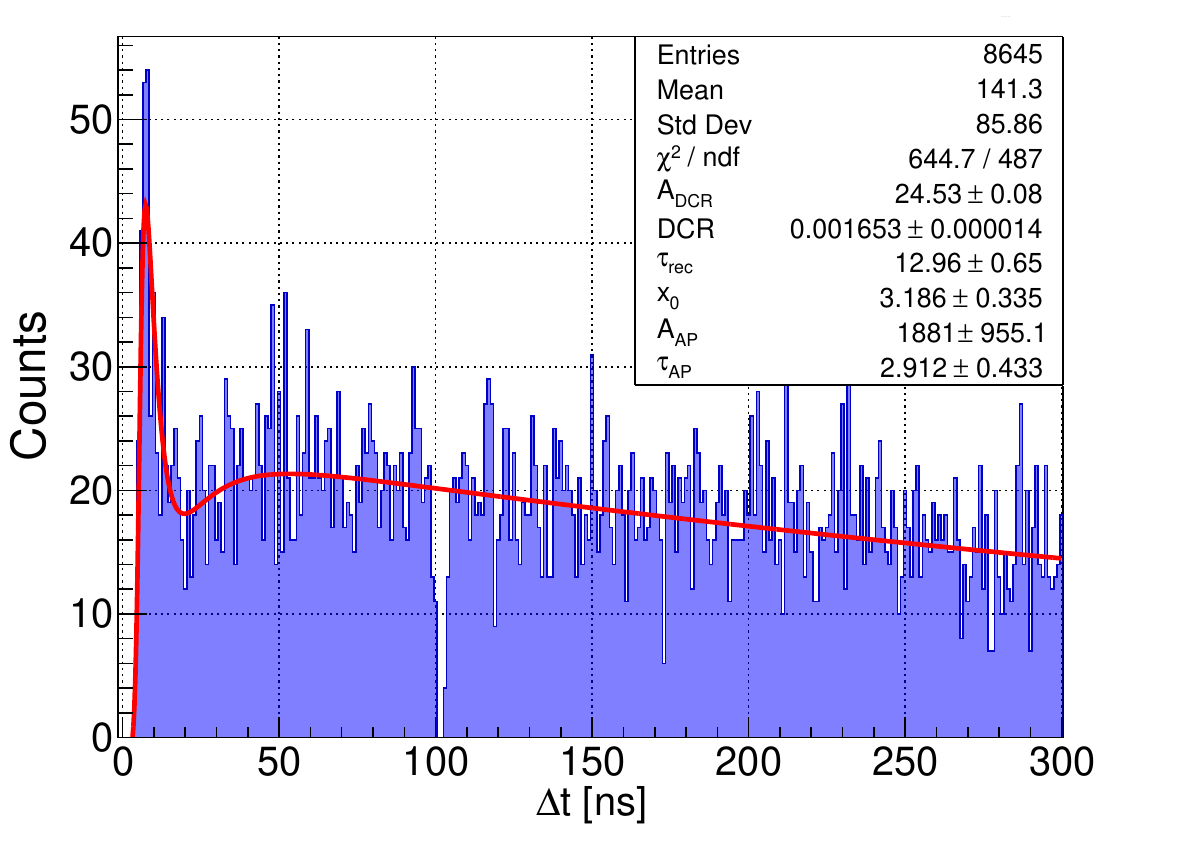}
        \caption{First secondary event separation.}
        \label{fig:first_secondary}
    \end{subfigure}
    \caption{Time-interval analysis. (a) Temporal distribution across waveforms showing DCR, veto, and signal regions. (b) Time separation $\Delta t$ (blue) fitted with the corrected afterpulsing function (red) from Sec.~\ref{sec:AP_extraction}; the gap at $\approx 100$~ns corresponds to laser reflection rejection.}
    \label{fig:dcr_deltat}
\end{figure}

The Dark Count Rate (DCR) is determined in a "dark" gate ($t_{gate}$) preceding the laser trigger: $DCR = N_{p} / (N_{w} \cdot t_{gate})$, where $N_{p}$ is the total pulse count across $N_{w}$ waveforms. Primary laser-induced pulses are tagged and preselected (see Fig.~\ref{fig:found_pulses}) based on three criteria: (i) a peak timestamp within $313 \pm 1.5$~ns, (ii) a normalized amplitude of $1 \pm 0.15$ relative to the single-photoelectron reference, and (iii) a 90~ns "veto" window preceding the trigger to ensure full cell recovery. For waveforms containing exactly one such primary discharge, the time interval $\Delta t$ to the first subsequent secondary pulse within the signal region is histogrammed for analysis (see Fig.~\ref{fig:first_secondary}).

\subsection{Extraction of Afterpulsing Parameters}
\label{sec:AP_extraction}
Afterpulsing parameters are extracted by fitting the time-interval distribution with the model:
\begin{equation}
N_{sec}(\Delta t) = \left[ N_{DC}(\Delta t) + N_{AP}(\Delta t) \right] \cdot R(\Delta t) \cdot \epsilon(\Delta t),
\label{eq:N_sec}
\end{equation}
where $N_{DC}(\Delta t) = A_{DCR}\cdot e^{-DCR \cdot\Delta t}$ accounts for the background and $N_{AP}(\Delta t) = A_{AP}\cdot e^{-\Delta t / \tau_{AP}}$ models the secondary discharges with de-trapping time constant $\tau_{AP}$. The normalization constants $A_{DCR}$ and $A_{AP}$ represent contributions at $\Delta t = 0$. Sensor constraints are incorporated via the composite recovery function $R(\Delta t) = \prod_{i=1}^{2} R_i(\Delta t)$, with $R_i(\Delta t)= 1 - \exp(-(\Delta t - x_{0,i}) / \tau_{i})$, accounting for Geiger discharge probability and charge recovery. The relationship between recovery constants $\tau_{i}$ follows \cite{GARUTTI2021165853}. The model is further corrected by the algorithm efficiency $\epsilon(\Delta t)$ (Sec.~\ref{sec:pulse_finder}).

The AP probability $P_{AP}$ is determined by integrating the corrected function $f_{AP}(\Delta t) = N_{AP}(\Delta t)\cdot R(\Delta t)$ such that $P_{AP} = \frac{1}{N_{laser}} \int_{t_{min}}^{t_{max}} f_{AP}(t)dt$. For a rigorous treatment, the first secondary event distribution must account for the survival probability:
\begin{equation}
N_{sec}(\Delta t) = \lambda(\Delta t)\cdot R(\Delta t)\cdot \epsilon(\Delta t)\cdot \exp\left( -\int_{0}^{\Delta t} \lambda(t') dt' \right),
\end{equation}
where $\lambda(\Delta t)$ is the instantaneous rate. However, since $P_{AP} < 5\%$ for $\Delta U < 5$~V, the exponential survival term approaches unity, making the simplified model in Eq.~\ref{eq:N_sec} a sufficient approximation for this low-occupancy regime.
\section{Simulations and validations}
\label{sec:simulations}

\begin{figure}[htbp]
     \centering
     \begin{subfigure}[t]{0.38\textwidth}
         \centering
         \includegraphics[width=\textwidth, trim={0 0 0 0.5cm}, clip]{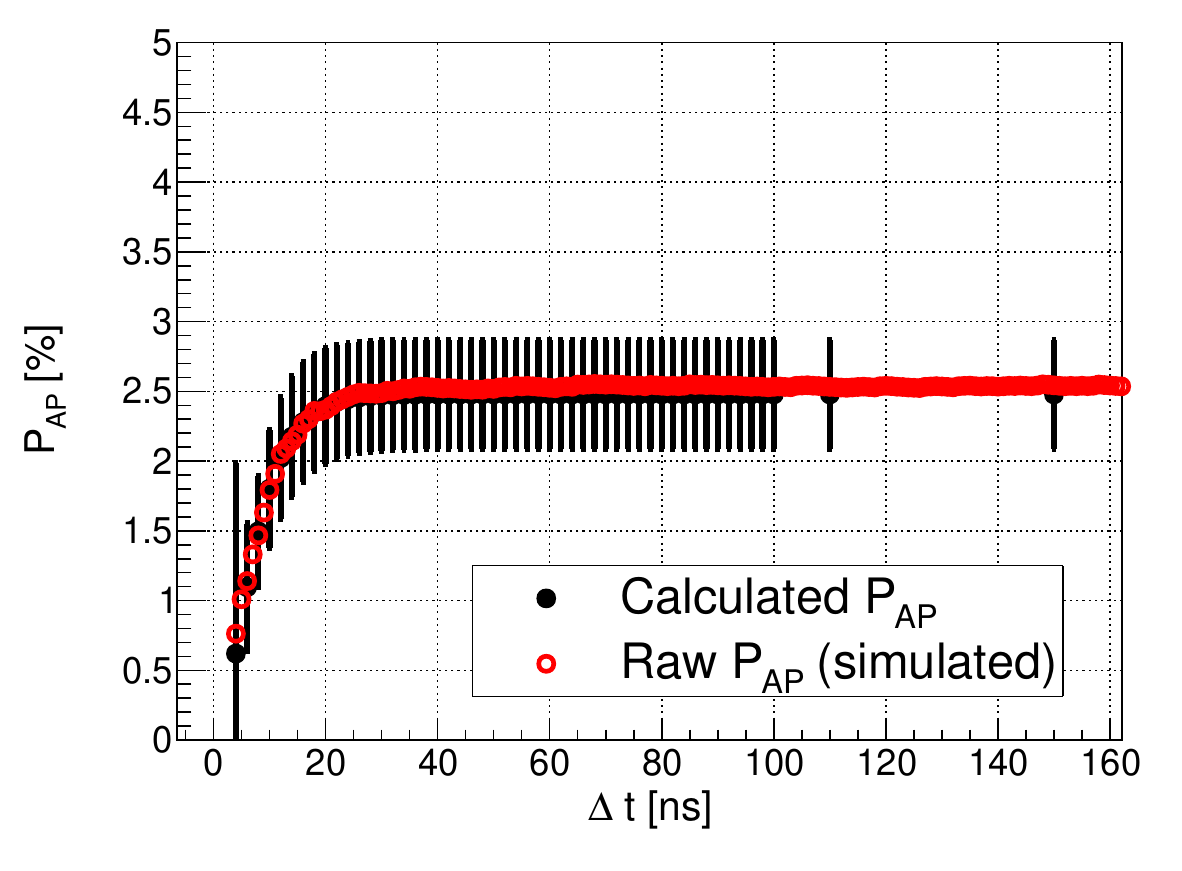}
         \caption{Validation against simulation.}
         \label{fig:pap_validation}
     \end{subfigure}
     \hfill
     \begin{subfigure}[t]{0.29\textwidth}
         \centering
         \includegraphics[width=\textwidth, trim={0 0 0 1cm}, clip]{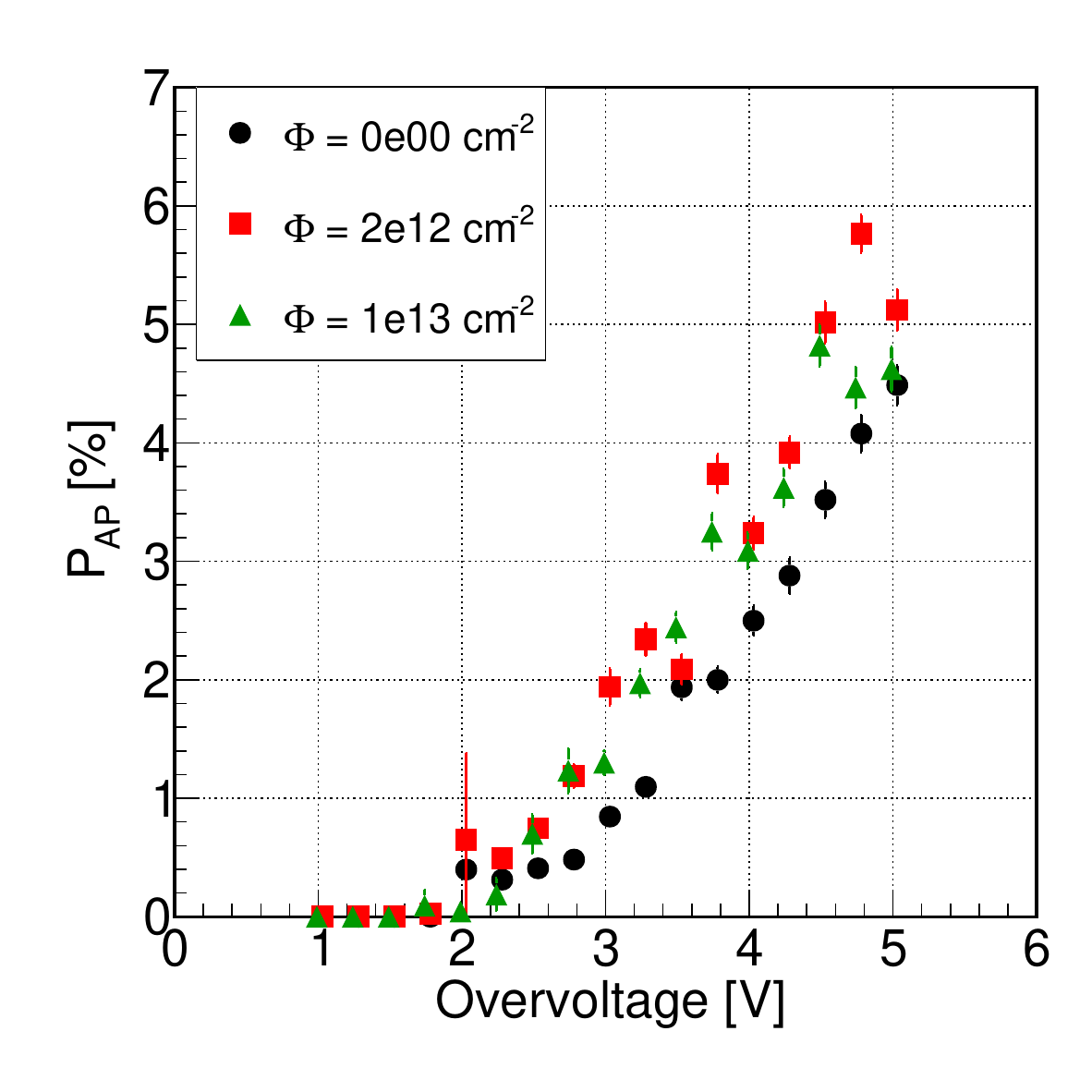}
         \caption{$P_{AP}$ vs. overvoltage.}
         \label{fig:AP_results}
     \end{subfigure}
     \hfill
     \begin{subfigure}[t]{0.29\textwidth}
         \centering
         \includegraphics[width=\textwidth, trim={0 0 0 1cm}, clip]{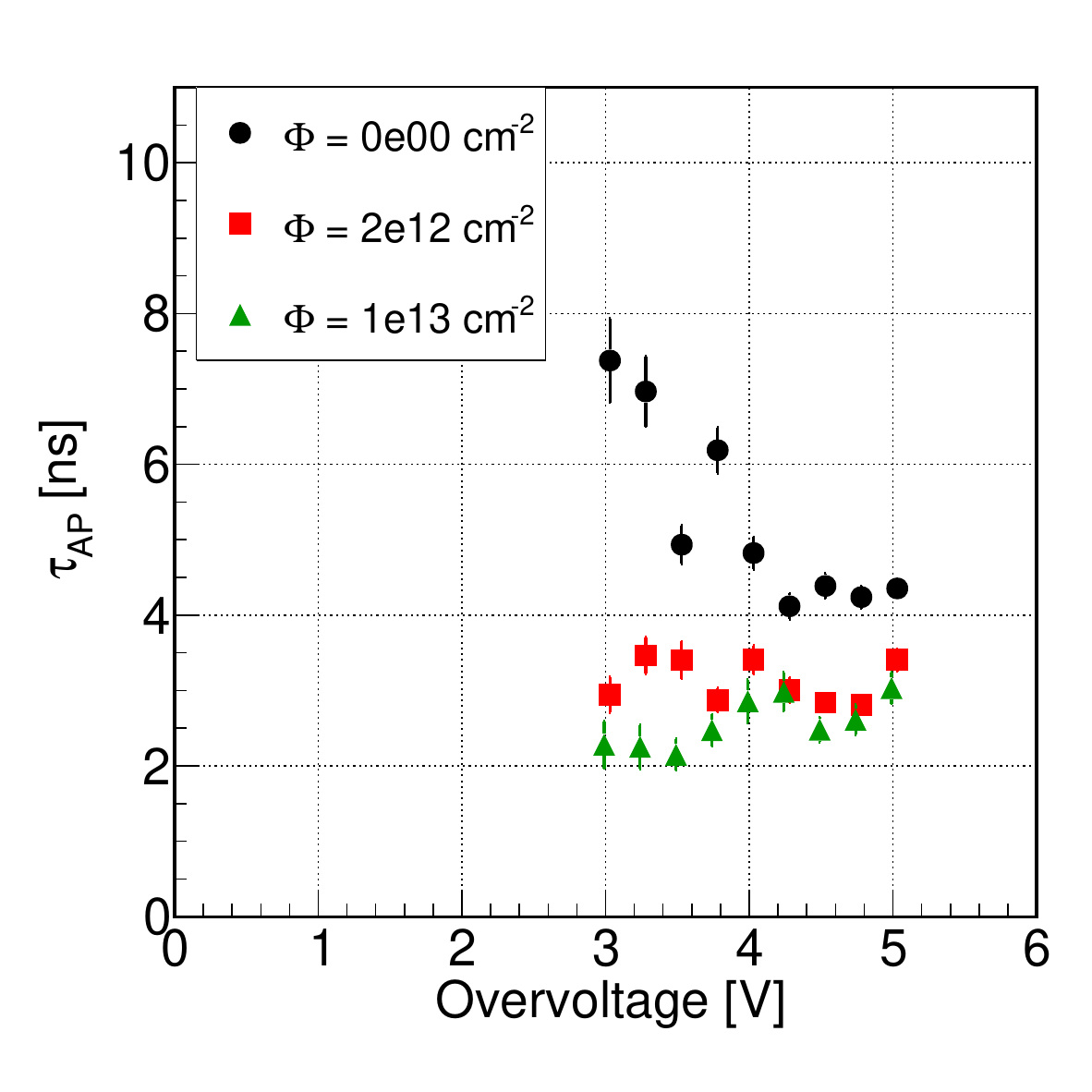}
         \caption{$\tau_{AP}$ vs. overvoltage.}
         \label{fig:tau_AP_results}
     \end{subfigure}
     \caption{Afterpulsing characterization and validation. (a) Comparison of $\Delta t$ extraction (black) with simulated truth (red) at $\Phi = 10^{13}~$cm$^{-2}$ and $\Delta U = 3.5~$V. In (b, c), non-irradiated reference sample is shown in black; irradiated samples are shown in red and green for $\Phi =2\cdot10^{12}$cm$^{-2}$ and $\Phi =1\cdot10^{13}~$cm$^{-2}$, respectively.}
     \label{fig:result_plots}
\end{figure}

The robustness of the proposed analysis was verified through a single-cell Monte Carlo simulation framework \cite{GARUTTI2021165853}. These simulations (Fig.~\ref{fig:raw_wf_sim}) incorporated recursive afterpulsing and electronic noise profiles derived from experimental data, while excluding optical crosstalk. Trials conducted for all three devices across an over-voltage range of $\Delta U_{OV} = 1\text{--}4$~V demonstrated excellent agreement between the simulated afterpulsing probability ($P_{AP}$) and the values extracted via the time-intervals method (Fig.~\ref{fig:pap_validation}). 

\section{Results and discussion}
\label{sec:results}

The characterization of afterpulsing parameters yields a total probability $P_{AP} < 6\%$ for overvoltages $\Delta U < 5$~V (Fig.~\ref{fig:AP_results}), with the integrated $P_{AP}$ saturating for $\Delta t > 30$~ns. The extraction also reveals a fast de-trapping time constant $\tau_{AP} < 10$~ns within the bias range $\Delta U = 3\text{--}5$~V (Fig.~\ref{fig:tau_AP_results}). Within the $600$~ns measurement window, neither parameter scales with irradiation fluence across the tested range, although a difference is visible between the non-irradiated reference and the irradiated devices: most notably a roughly twofold increase of $P_{AP}$ at $3.8$~V overvoltage and a several-fold drop of $\tau_{AP}$ at low overvoltages. Among the irradiated samples, variations between $2\cdot10^{12}$ and $1\cdot10^{13}$~cm$^{-2}$ remain within or close to the experimental uncertainties. The fact that $\tau_{AP}$ stays in the few-nanosecond range for all irradiated devices points to very shallow defects or, possibly, optically-induced delayed cross-talk as the dominant contribution, rather than radiation-induced deep traps. The $600$~ns window is, however, insensitive to slower trapping on the microsecond scale or above, which may exhibit a different fluence dependence.

A comprehensive understanding requires extending the measurement window to the microsecond scale to identify potential long-lived trapping states. Temperature-dependence studies are also essential to determine activation energies and definitively distinguish between lattice defects and delayed cross-talk. These combined efforts will provide a more complete picture of afterpulsing dynamics in irradiated SiPMs.




\bibliographystyle{JHEP}
\bibliography{biblio.bib}

\end{document}